\documentclass[12pt]{article}
\usepackage{amsmath,amssymb}
\begin{document}
\title{\huge \bf Octonionic Geometry}
\author{{\bf Merab Gogberashvili} \\
Andronikashvili Institute of Physics \\
6 Tamarashvili Str.,Tbilisi 380077, Georgia \\
{\sl E-mail: gogber@hotmail.com }} \maketitle
\begin{abstract}
We extend vector formalism by including it in the algebra of split
octonions, which we treat as the universal algebra to describe
physical signals. The new geometrical interpretation of the
products of octonionic basis units is presented. Eight real
parameters of octonions are interpreted as the space-time
coordinates, momentum and energy. In our approach the two
fundamental constants, $c$ and $\hbar$, have the geometrical
meaning and appear from the condition of positive definiteness of
the octonion norm. We connect the property of non-associativity
with the time irreversibility and fundamental probabilities in
physics.

\end{abstract}
\medskip

%\newpage
%%%%%%%%%%%%%%%%%%%%%%%%%%%%%%%%%%%%%%%%%%%%%%%%%%%%%%%%%%%%%%%%%

\section{Introduction}

Usually in physics the geometry is thought to be objective without
any connection to the way it is observed. However, the properties
of space-time (such as dimension, distances, etc.) are just
reflections of the symmetries of physical signals we receive.
Analyzing signals our brain can operate only via the classical
associations and numbers. Algebras closely relates to the physical
measurements, which consists not only with the actual receiving of
signals, but with perceiving this signals to form the data as
well. So we can introduce some kind of anthrophic principle in
mathematics, which means that investigating the algebras we study
the way our brain abstracts physical observations. For example,
algebra of rational numbers expresses our experience that there
exist some classical objects with which we can exchange signals
and these objects do not change much during 'algebraic'
operations. Any observable quantity, which our brain could extract
from a single measurement, is a real number, the norm (distance)
formed by the multiplication of the numbers corresponding to the
direct and reflected signals. In general one-way signal can be
expressed with any kind of number, important is to have the real
norm. Introduction of some distance (norm) always means some
comparison of two physical objects using one of them as an etalon
(for example, simple counting). In the algebraic language all
these features of our way of thinking mean that to perceive the
real world our brain uses division algebras with the unit element
over the field of real numbers.

Besides of the usual algebra of real numbers there are, according
to the Hurwitz theorem, three unique division algebras, the
algebra of complex numbers, quaternions and octonions \cite{Sc}.
Essential feature of all normed composition  algebras is the
existence of a real unit element and a different number of
adjoined hyper-complex units. The square of the unit element is
always positive while the squares of the hyper-complex units can
be negative as well. In applications of division algebras mainly
the elements with the negative square (similar to the complex unit
$i$) are used. In this case norm of the algebra is positively
defined. Using the vector-like elements (with the positive square)
in division algebras leads to so-called split algebras having an
equal number of terms with the positive and negative signs in the
definition of their norms.

Because of importance of 'imaginary' and vector-like elements of
algebras we want to recall the history of their introduction
\cite{Sa}.

Word 'vector' was introduced by William Hamilton for the pure
imaginary part of quaternions, discovered by him in 1843. He did
not give geometrical meaning to the scalar part of a quaternion
but kept it to have division property of algebra. Later Hamilton
used also the word 'versors' (meaning 'rotators') for three
quaternionic basis units with the negative square, since he
understood that such elements could be interpreted as representing
rotation of vectors, instead of something that corresponds to a
straight line in space.

The algebra of Euclidean vectors was developed in 1880is by Gibbs
and Heaviside from Hamilton's quaternions when they tried to
rewrite Maxwell's quaternionic equations in a more convenient
form. They removed scalar part of quaternions and kept Hamilton's
term 'vector' representing a pure quaternion. Since vectors have
their origin in physical problems the definition of the products
of vectors was obtained from the way in which such products occur
in physical applications. Instead of the whole quaternion product
(which has a scalar and a 'vector' part), Gibbs and Heaviside
defined two different types of vector multiplications, the scalar
and vector products. They also changed the sign from minus to plus
in the scalar product, resulting to positive squares of basis
units. This changing corresponds to a shift of interpretation of
basis units from 'versors' to unit vectors.

By revising the process of addition of physical vectors it was
found that using the three unit orthogonal basis elements $e_n$
($n = 1,2,3$) any vector could be decomposed into components. One
can introduce also conjugated (reflected) vectors
$\widetilde{e_n}$, which differ from $e_n$ by the sign. The
properties of scalar and vector products are encoded in the
following algebra of the unit orthogonal basis elements
\begin{equation} \label{e-vector}
 e_n^2 = 1~,~~~~ \widetilde{e_n} = - e_n~,~~~~
 e_ne_m = - e_me_n = \varepsilon_{nmk}e^k
 ~,~~~~~~~~~~n,m = 1,2,3
\end{equation}
where $\varepsilon_{nmk}$ is fully anti-symmetric tensor. Or vice
versa, we can say that postulating the algebra (\ref{e-vector})
one can recover the ordinary multiplication laws of physical
vectors. There is ambiguity in choosing of the sign of
$\varepsilon_{nmk}$ in (\ref{e-vector}) connected with the
exitance of the left-handed and right-handed coordinate systems.
Note that anti-commutation of the basis units in (\ref{e-vector})
is the result of the definition of vector product, and not an
essential property of the vector basis elements $e_n$ themselves.

Introduction of vector algebra in physics was successful, however
because of removing of the scalar part of quaternions the division
operation is not defined for vectors. There was lost also the
property of 'versors', that they are rotation generators and
expresses not only the final state achieved after a rotation, but
the direction in which this rotation has been performed. It is
this direction of rotation that the standard matrix representation
of the rotation group fails to give.

It is known that Hamilton's pure quaternions are not equivalent to
vectors \cite{SiMa}. For ordinary Euclidean vectors, basis
elements are three orthogonal unit polar vectors, while for
Hamilton's quaternions they are unit 'versors' (imaginary units)
and with respect to coordinate transformations behave similar to
axial vectors. There are differences also in the proprieties of
their products. The product of quaternions is associative, while
both types of vector products are not. So the division algebra,
which includes the formalism of Euclidean 3-vectors, should be
wider then the algebra of quaternions.

In this paper we present extension of vector algebra
(\ref{e-vector}) by embedding it in the algebra of
split-octonions. In this sense our approach is a generalization of
the well-developed formalism of geometric algebras \cite{He, Ba}.
Split octonions contain exactly three vector-like orthogonal
elements needed to describe special dimensions \cite{Go}.

%%%%%%%%%%%%%%%%%%%%%%%%%%%%%%%%%%%%%%%%%%%%%%%%%%%%%%%%%%%%%%%%%%

\section{Octonionic Intervals}

Since their discovery in 1844-1845 by Graves and Cayley there were
various attempts to find appropriate uses for octonions in physics
\cite{Oct}. Recently the assumption that our Universe is made of
pairs of octonions become an important idea in string theory also
\cite{string}. As distinct from string models in present paper we
want to apply octonions to describe space-time geometry and not
only internal spaces.

We want to describe the physical signal $s$ by a 8-dimensional
number, the element of octonionic algebra,
\begin{equation} \label{O}
s = ct  + x_nJ^n + \hbar \lambda_nj^n +
c\hbar\omega I~, ~~~~~~~~~~~~ n = 1, 2, 3
\end{equation}
where by the repeated indexes summing is considered as in standard
tensor calculus. The eight scalar parameters in (\ref{O}) we treat
as the time $t$, special coordinates $x^n$, quantities $\lambda^n$
with the dimension of the momentum$^{-1}$ and $\omega$ with the
dimension of energy$^{-1}$. In (\ref{O}) two fundamental constants
of physics, the velocity of light $c$ and Plank's constant
$\hbar$, are presented also \cite{Go}.

The orthogonal basis of split-octonions (\ref{O}) is formed by the
unit scalar element (which we denote by $1$) and by three
different types (totally seven) of orthogonal hyper-complex units:
the three vector-like elements $J_n$, three 'versor'-like units
$j_n$ and one pseudo-scalar $I$. Similar quantity, represented as
a sum of the elements with the different properties, is called
para-vector in Clifford algebras \cite{Clifford}.

The squares of octonionic basis units
\begin{equation} \label{JjI}
J_n^2=1~,~~~~~ j_n^2=-1~, ~~~~~ I^2=1~, ~~~~~~~~~~n = 1, 2, 3
\end{equation}
are always inner products resulting unit element. However,
multiplication of different basis units should be defined as the
skew products
\begin{eqnarray} \label{OA}
J_nJ_m = - J_mJ_n ~, ~~~~j_nj_m = - j_mj_n ~,~~~~~ \ J_nj_m = -
j_mJ_n ~, \nonumber \\
J_nI = -IJ_n ~, ~~~~ j_nI = - Ij_n ~. ~~~~~~~~~~~~~~~~(n \ne m)
\end{eqnarray}
As distinct from the geometric algebra approaches \cite{He, Ba},
here we do not need to introduce different types of brackets for
inner and outer products, similar to the algebra of Euclidean
vectors (\ref{e-vector}).

To generate complete 8-dimensional basis of split-octonions ( $1,
J^n, j^n$ and $I$) the multiplication and distribution laws of
only three vector-like elements $J_n$ are needed ($2^n = 8$). Then
our imagination about 3-dimensional character of the space can be
the result of existing of only three vector-like basis units. The
fundamental basis elements $J_n$ geometrically can be presented as
the unit orthogonal Euclidean vectors
\begin{equation}
J_1 = \setlength{\unitlength}{0.3mm}
\begin{picture}(20, 20)\thicklines
\put(0,5){\vector(1,0){20}}
\end{picture}~,
~~~~~J_2 = \setlength{\unitlength}{0.3mm}
\begin{picture}(20, 20)\thicklines
\put(10,0){\vector(0,1){17}}
\end{picture} ,
~~~~~J_3 = \setlength{\unitlength}{0.3mm}
\begin{picture}(20, 20)\thicklines
\put(12,10){\vector(-1,-1){11.8}}
\end{picture} ,
\end{equation}
directed towards the positive directions of $x, y$ and $z$ axis
respectively. As for the vector algebra (\ref{e-vector}),
conjugated elements $\widetilde{J_n}$ can be understood as the
reflected vectors
\begin{equation}
\widetilde{J_1}= -J_1 = \setlength{\unitlength}{0.3mm}
\begin{picture}(20, 20)\thicklines
\put(20,5){\vector(-1,0){20}}
\end{picture}~,
~~~~~\widetilde{J_2} =-J_2 = \setlength{\unitlength}{0.3mm}
\begin{picture}(20, 20)\thicklines
\put(10,15){\vector(0,-1){17}}
\end{picture} ,
~~~~~\widetilde{J_3}=-J_3 = \setlength{\unitlength}{0.3mm}
\begin{picture}(20, 20)\thicklines
\put(0,0){\vector(1,1){11.8}}
\end{picture} .
\end{equation}

The left multiplication of any hyper-complex element on the
vector-like element $J_n$ we visualize geometrically as sweeping
of this element along $J_n$. In this picture square of $J_n$ means
sweeping of $J_n$ along the other $J_n$ taking its origin to the
end value (which is equal to $1$)
\begin{eqnarray}
J_nJ_n = J_n^2 =
\setlength{\unitlength}{0.3mm}
\begin{picture}(20, 20)\thicklines
\put(0,5){\vector(1,0){20}}
\end{picture}
\times \setlength{\unitlength}{0.3mm}
\begin{picture}(20, 20)\thicklines
\put(0,5){\vector(1,0){20}}
\end{picture}
= ~_0\dashrightarrow_{~1} ~= 1 ~, \nonumber \\
J_n\widetilde{J_n} = - J_n^2 =
\setlength{\unitlength}{0.3mm}
\begin{picture}(20, 20)\thicklines
\put(0,5){\vector(1,0){20}}
\end{picture}
\times \setlength{\unitlength}{0.3mm}
\begin{picture}(20, 20)\thicklines
\put(20,5){\vector(-1,0){20}}
\end{picture}
= ~_{-1}\dashleftarrow_{~0} ~= -1 ~.
\end{eqnarray}

Another class of basis elements $j^n$ in (\ref{O}) (which are dual
to $J^n$) is defined as the skew product of two fundamental basis
units $J^n$
\begin{equation} \label{j}
j_n = \frac{1}{2}\epsilon _{nmk}J^mJ^k~. ~~~~~~~~~~~~~~~~~ m,n,k = 1,
2, 3
\end{equation}
The elements $j_n$ are neither a scalar nor a vector, we interpret
them as bi-vectors, similar to \cite{He, Ba}. Since $j_n^2 = -1$,
bi-vectors behave like pure imaginary objects. There are
differences, however. Here we have three anti-commuted bi-vectors
and also they anti-commute with the vector-like objects ($j_nJ_m =
-J_mj_n$). We do not get behavior like this with complex numbers
alone. The feature which imitates complexity and results in
"imaginary" properties of $j_n$ is the definition of the products
of $J_n$. So the objects $j_n$, similar to Hamilton's quaternionic
units, are 'versors' and can be used to represent rotations. As
for the vector products (\ref{e-vector}), we have ambiguity in
choosing the sign of $\epsilon _{nmk}$ in the relation (\ref{j}).
In quaternionic and vector algebras this ambiguity is often
understood as connected with the left-handed and right-handed
coordinate systems. In our case elements $j_n$ encodes the notion
of an oriented plane without relying on the notion of a vector
perpendicular to it. In the relation (\ref{j}) we also choose
positive sign, however, when considering products of octonions
corresponding to the different signals, the ambiguity still
remains and in physical applications could give two-value
wave-functions, corresponding to the introduction of spin. Also
the elements $j_n$ are useful to define momentum operators similar
as done in quaternionic quantum mechanics \cite{So}. This is one
of the justifications of appearance of Plank's constant in the
definition (\ref{O}).

We want to visualize $j_n$ geometrically as the orientated planes
obtained by the sweeping of the second fundamental vector in the
definition (\ref{j}) along the first. For example,
\begin{equation}\label{j3}
j_3 = J_1J_2 =
\setlength{\unitlength}{0.3mm}
\begin{picture}(20, 20)\thicklines
\put(0,5){\vector(1,0){20}}
\end{picture}
~\times \setlength{\unitlength}{0.3mm}
\begin{picture}(20, 20)\thicklines
\put(10,0){\vector(0,1){17}}
\end{picture}
=~
\begin{picture}(20, 20)\thicklines
\put(0,0){\vector(1,0){20}} \put(0,3){\vector(0,1){17}} \thinlines
\put(5,10){\vector(1,0){10}}
\end{picture}
~=~
\begin{picture}(20, 20)
\thicklines \put(0,5){\line(1,0){20}} \put(20,5){\vector(0,1){10}}
\end{picture} ~.
\end{equation}
Changing the order of vectors in (\ref{j3}) reverses the orientation of the
plane
\begin{equation}\label{-j3}
J_2J_1 =-j_3 = \setlength{\unitlength}{0.3mm}
\begin{picture}(20, 20)\thicklines
\put(10,0){\vector(0,1){17}}
\end{picture}
\times ~\setlength{\unitlength}{0.3mm}
\begin{picture}(20, 20)\thicklines
\put(0,5){\vector(1,0){20}}
\end{picture}
~=~
\begin{picture}(20, 20)\thicklines
\put(3,0){\vector(1,0){17}} \put(0,0){\vector(0,1){17}} \thinlines
\put(10,5){\vector(0,1){10}}
\end{picture}
~=
\begin{picture}(20, 20)
\thicklines \put(5,0){\line(0,1){17}} \put(5,17){\vector(1,0){10}}
\end{picture}.
\end{equation}
Analogously for other two bi-vectors $j_1$ and $j_2$ we find
\begin{eqnarray} \label{j2j1}
J_1J_3 = -j_2 = \setlength{\unitlength}{0.3mm}
\begin{picture}(20, 20)\thicklines
\put(0,5){\vector(1,0){18}}
\end{picture}
\times
\begin{picture}(20, 20)\thicklines
\put(15,10){\vector(-1,-1){11.8}}
\end{picture}
=~
\begin{picture}(20, 20)\thicklines
\put(8,8){\vector(-1,-1){11.8}} \put(5,10){\vector(1,0){17}}
\thinlines \put(8,2){\vector(1,0){12}}
\end{picture}
~=
\begin{picture}(20, 20)
\thicklines \put(0,10){\line(1,0){20}}
\put(20,10){\vector(-1,-1){11.8}}
\end{picture} , \nonumber \\
J_2J_3 = j_1 =
\setlength{\unitlength}{0.3mm}
\begin{picture}(20, 20)\thicklines
\put(10,0){\vector(0,1){17}}
\end{picture}
\times \setlength{\unitlength}{0.3mm}
\begin{picture}(20, 20)\thicklines
\put(15,10){\vector(-1,-1){11.8}}
\end{picture}
=
\begin{picture}(20, 20)\thicklines
\put(12,5){\vector(-1,-1){11.8}} \put(15,3){\vector(0,1){17}}
\thinlines \put(5,3){\vector(0,1){12}}
\end{picture}
=
\begin{picture}(20, 20)
\thicklines \put(15,0){\line(0,1){17}}
\put(15,17){\vector(-1,-1){11.8}}
\end{picture}.
\end{eqnarray}
In our geometric approach sub-algebra of products of the only two
vector-like units $J_n$ can be easily found from the rotation of
the figures (\ref{j3}), (\ref{-j3}) and (\ref{j2j1}). For example,
the commutation laws
\begin{equation}
J_1J_2 = J_2\widetilde{J_1}= \widetilde{J_1}\widetilde{J_2} =
\widetilde{J_2}J_1~,
\end{equation}
have the geometrical interpretation as the rotations of the figure
(\ref{j3}) in the $(x-y)$-plane
\begin{equation}
\setlength{\unitlength}{0.3mm}
\begin{picture}(20, 20)
\thicklines \put(0,5){\line(1,0){20}} \put(20,5){\vector(0,1){10}}
\end{picture}~
=
\begin{picture}(20, 20)\thicklines
\put(10,0){\line(0,1){17}} \put(10,17){\vector(-1,0){10}}
\end{picture}
=~
\begin{picture}(20, 20)\thicklines
\put(20,10){\line(-1,0){20}} \put(0,10){\vector(0,-1){10}}
\end{picture}~
=
\begin{picture}(20, 20)\thicklines
\put(5,17){\line(0,-1){17}} \put(5,0){\vector(1,0){10}}
\end{picture}.
\end{equation}
Analogously the laws
\begin{equation}
J_2J_1= J_1\widetilde{J_2} = \widetilde{J_2}\widetilde{J_1}=
\widetilde{J_1}J_2~,
\end{equation}
correspond to the rotations of the oppositely orientated figure
(\ref{-j3}),
\begin{equation}
\setlength{\unitlength}{0.3mm}
\begin{picture}(20, 20)\thicklines
\put(10,0){\line(0,1){17}} \put(10,17){\vector(1,0){10}}
\end{picture}
=~
\begin{picture}(20, 20)\thicklines
\put(20,10){\line(-1,0){20}} \put(20,10){\vector(0,-1){10}}
\end{picture}~
=
\begin{picture}(20, 20)\thicklines
\put(15,17){\line(0,-1){17}} \put(15,0){\vector(-1,0){10}}
\end{picture}
=~
\begin{picture}(20, 20)
\thicklines \put(0,5){\line(1,0){20}} \put(0,5){\vector(0,1){10}}
\end{picture}~.
\end{equation}
With the similar rotations of (\ref{j2j1}) in the $(x-z)$- and
$(y-z)$-planes we recover binary products of all three vector-like
basis units $J_n$ and their conjugates $\widetilde{J_n}$.

Non-associativity of octonions, which results in non-equivalence
of left and right products for the expression containing more then
two basis units $J_n$, physically can be interpreted as the
causality. To the direction from the past to the future we
correspond one orientation of multiplication, for example the left
product.

To find the expression of products of $j_n$ on $J_m$ we can use
the property of alternativity of the octonions. In the language of
basis elements alternativity means that in the expressions of the
multiplication of several basis units, where only two different
fundamental basis elements are involved, the order of products is
arbitrary. Physically this means that for the physical processes
taking place in a single plane time is reversible. So products of
$J^n$ and $j^m$ (when $n\ne m$) can be defined as
\begin{equation} \label{Jj}
J_n j_m = - \epsilon_{nmk}J^k ~.
\end{equation}
The products of $J_n$ by $j_m$  we visualize geometrically as
rotations of the vectors $J_n$. For example, left product of $j^3
$ on $J^2$ (or $J_1$) is the operation meaning that $j^3$ applied
to $J^2$ (or $J_1$) rotates it clockwise by $\pi/2$ in the
$(x-y)$-plane, resulting to $J^1$ (or $-J_2$)
\begin{eqnarray}\label{j1J1}
j_3J_2=(J_1J_2)J_2 = J_1 =\setlength{\unitlength}{0.3mm}
\begin{picture}(20, 20)
\thicklines \put(0,5){\line(1,0){20}} \put(20,5){\vector(0,1){10}}
\end{picture}
~\times
\begin{picture}(15, 20)\thicklines
\put(7,0){\vector(0,1){17}}
\end{picture}
=\frac{\pi}{2}\curvearrowright
\begin{picture}(15, 20)\thicklines
\put(7,0){\vector(0,1){17}}
\end{picture}
= \setlength{\unitlength}{0.3mm}
\begin{picture}(20, 20)\thicklines
\put(0,5){\vector(1,0){20}}
\end{picture}
~,\nonumber \\
j_3J_1=(J_1J_2)J_1 = -J_2 =\setlength{\unitlength}{0.3mm}
\begin{picture}(20, 20)
\thicklines \put(0,5){\line(1,0){20}} \put(20,5){\vector(0,1){10}}
\end{picture}
~\times~
\begin{picture}(20, 20)\thicklines
\put(0,5){\vector(1,0){20}}
\end{picture}
~=\frac{\pi}{2}\curvearrowright
\begin{picture}(20, 20)\thicklines
\put(0,5){\vector(1,0){20}}
\end{picture}
= \setlength{\unitlength}{0.3mm}
\begin{picture}(15, 20)\thicklines
\put(7,17){\vector(0,-1){17}}
\end{picture}.
\end{eqnarray}

Right products of $J_n$ by $j_m$ ($n\ne m$) can be considered as
anti-clockwise rotations, or alternatively as the left sweepings
of the oriented planes $j_n$ along the vectors $J_m$. For example,
to the products reverse to (\ref{j1J1}) correspond
\begin{eqnarray} \label{J1j1}
J_1j_3 = J_1(J_1J_2) = ~\setlength{\unitlength}{0.3mm}
\begin{picture}(20, 20)\thicklines
\put(0,5){\vector(1,0){20}}
\end{picture}
~\times~
\begin{picture}(20, 20)
\thicklines \put(0,5){\line(1,0){20}} \put(20,5){\vector(0,1){10}}
\end{picture}
~= \frac{\pi}{2}\curvearrowleft
\begin{picture}(20, 20)\thicklines
\put(0,5){\vector(1,0){20}}
\end{picture}
 = \setlength{\unitlength}{0.3mm}
\begin{picture}(15, 20)\thicklines
\put(7,0){\vector(0,1){17}}
\end{picture}
, \nonumber \\
J_2j_3 = J_2(J_1J_2) = \setlength{\unitlength}{0.3mm}
\begin{picture}(20, 20)\thicklines
\put(10,0){\vector(0,1){17}}
\end{picture}
\times~
\begin{picture}(20, 20)
\thicklines \put(0,5){\line(1,0){20}} \put(20,5){\vector(0,1){10}}
\end{picture}
~=\frac{\pi}{2}\curvearrowleft
\begin{picture}(15, 20)\thicklines
\put(7,0){\vector(0,1){17}}
\end{picture}
=
 \setlength{\unitlength}{0.3mm}
\begin{picture}(20, 20)\thicklines
\put(20,5){\vector(-1,0){20}}
\end{picture}
~.
\end{eqnarray}
Pictures similar to (\ref{j1J1}) and (\ref{J1j1}) we can paint
also in the $(x-z)$- and $(y-z)$-planes.

The seventh octonionic basic element $I$ (dual to the scalar unit
$1$) we treat as oriented 3-vector, maximal multi-vector generated
by three elements $J^n$. We define $I$ as the left product of all
fundamental vectors $J_n$,
\begin{equation} \label{I}
I = J_nj_n = - j_nJ_n  ~.
\end{equation}
There is no summing in this formula. From (\ref{I}) we see that
$I$ has the three equivalent representations
\begin{equation} \label{I'}
I = J_1j_1 =  J_2j_2 = J_3j_3~,
\end{equation}
which can be understood as the expression of the volume
invariance. The element $I$ can be visualized as 3-dimensional
oriented cube obtained by the sweeping of oriented plane $j_n$
along the vector $J_n$. For example,
\begin{equation} \label{cube}
I = J_1j_1 = J_1(J_2J_3) = \setlength{\unitlength}{0.3mm}
\begin{picture}(20, 20)\thicklines
\put(0,5){\vector(1,0){20}}
\end{picture}
~\times
\begin{picture}(20, 20)
\thicklines \put(12,0){\line(0,1){17}}
\put(12,17){\vector(-1,-1){11.8}}
\end{picture}
=~
\begin{picture}(20, 20)\thicklines
\put(0,0){\vector(1,0){20}} \put(5,3){\line(0,1){15}}
\put(5,17){\vector(-1,-1){11.8}} \thinlines
\put(8,10){\vector(1,0){12}}
\end{picture}
~=
\begin{picture}(20, 20)\thicklines
\put(0,0){\line(1,0){16}} \put(16,0){\line(0,1){17}}
\put(16,17){\vector(-1,-1){11.8}}
\end{picture}~.
\end{equation}
The other two equivalent representations of $I$
\begin{eqnarray}
J_2(J_3J_1)= J_2j_2 =\setlength{\unitlength}{0.3mm}
\begin{picture}(20, 20)\thicklines
\put(10,0){\vector(0,1){17}}
\end{picture}
\times
\begin{picture}(20, 20)
\thicklines \put(0,0){\vector(1,0){17}}
\put(12,12){\line(-1,-1){11.8}}
\end{picture}
=
\begin{picture}(20, 20)\thicklines
\put(0,5){\vector(1,0){22}} \put(12,0){\line(0,1){17}}
\put(12,17){\line(-1,-1){11.8}}
\end{picture}~, \nonumber \\
J_3(J_1J_2) = J_3j_3 =\setlength{\unitlength}{0.3mm}
\begin{picture}(20, 20)\thicklines
\put(14,10){\vector(-1,-1){11.8}}
\end{picture}
\times
\begin{picture}(20, 20)\thicklines
\put(3,5){\line(1,0){17}} \put(20,5){\vector(0,1){10}}
\end{picture}
~=
\begin{picture}(20, 20)\thicklines
\put(0,0){\line(1,0){17}} \put(17,0){\vector(0,1){12}}
\put(12,12){\line(-1,-1){11.8}}
\end{picture}~,
\end{eqnarray}
correspond to the different rotations of the figure (\ref{cube}).

Oppositely orientated 3-cubes are received by the left products of
$j_n$ on $J_m$. When $n\ne m$ similar left products of $j_n$ on
$J_m$ we had interpreted as the rotation of $J_m$. However, when
$n=m$ these elements 'align', $j_n$ can't rotate $J_n$ and we
visualize this product as attaching of $J_n$ by $j_n$
\begin{eqnarray}
j_1J_1 = (J_2J_3)J_1 = \setlength{\unitlength}{0.3mm}
\begin{picture}(20, 20)
\thicklines \put(14,0){\line(0,1){17}}
\put(14,17){\vector(-1,-1){11.8}}
\end{picture}
\times
\begin{picture}(20, 20)\thicklines
\put(0,5){\vector(1,0){20}}
\end{picture}
=~
\begin{picture}(20, 20)\thicklines
\put(10,0){\line(1,0){12}} \put(10,0){\line(0,1){17}}
\put(10,17){\vector(-1,-1){11.8}}
\end{picture}~~, \nonumber \\
j_2J_2 = (J_3J_1)J_2 =~\setlength{\unitlength}{0.3mm}
\begin{picture}(20, 20)
\thicklines \put(0,0){\vector(1,0){17}}
\put(12,12){\line(-1,-1){11.8}}
\end{picture}
\times
\begin{picture}(20, 20)\thicklines
\put(10,0){\vector(0,1){17}}
\end{picture}
=
\begin{picture}(20, 20)\thicklines
\put(-3,-3){\vector(1,0){17}} \put(9,9){\line(0,1){15}}
\put(9,9){\line(-1,-1){11.8}}
\end{picture}~, \\
j_3J_3 = (J_1J_2)J_3 =\setlength{\unitlength}{0.3mm}
\begin{picture}(20, 20)\thicklines
\put(3,5){\line(1,0){17}} \put(20,5){\vector(0,1){10}}
\end{picture}
~\times
\begin{picture}(20, 20)\thicklines
\put(14,10){\vector(-1,-1){11.8}}
\end{picture}
=~
\begin{picture}(20, 20)
\thicklines \put(20,10){\vector(0,1){15}}
\put(8,10){\line(1,0){12}} \put(8,10){\line(-1,-1){11.8}}
\end{picture}~.\nonumber
\end{eqnarray}

The essential property of octonions, non-associativity, is the
direct result of volume invariance requirement (\ref{I'}). Indeed,
\begin{equation}\label{123}
J_1(J_2J_3) - (J_1J_2)J_3= J_1j_1 - j_3J_3 = 2I =
\setlength{\unitlength}{0.3mm}
\begin{picture}(20, 20)\thicklines
\put(0,0){\line(1,0){16}} \put(16,0){\line(0,1){17}}
\put(16,17){\vector(-1,-1){11.8}}
\end{picture}
-
\begin{picture}(20, 20)
\thicklines \put(20,10){\vector(0,1){15}}
\put(8,10){\line(1,0){12}} \put(8,10){\line(-1,-1){11.8}}
\end{picture}
=
\begin{picture}(20, 20)\thicklines
\put(0,0){\line(1,0){16}} \put(16,0){\line(0,1){14.5}}
\put(16,14){\vector(-1,-1){11.8}} \put(16,14){\vector(1,1){11.8}}
\end{picture}
\neq 0~.
\end{equation}
Non-associativity is another property which shows that the
octonionic fundamental basis units $J^n$ are similar to the
ordinary 3-vectors, having non-associative products also.

Adopting non-associativity (\ref{123}), at the same time we need
the rules to receive definite results for products of all seven
octonionic basis units and their conjugates to form octonionic
algebra. This requirement forces us to introduce special rule for
opening of brackets in product of the vector-like elements $J^n$
and the 'versor'-like elements $j^m$ with the seventh basis unit
$I$. The products of $J^n$ and $j^n$ with $I = (J_mj_m)$ when $n =
m$, because of alternativity, gives
\begin{equation} \label{JIjI}
J_nI = j_n ~, ~~~~~ j_n I = J_n ~.
\end{equation}
However, when $n \ne m$ these products contains all three
fundamental basis units. If we ignore orientated feature of the
products and will try to remove the brackets, then, because of
anti-associativity (\ref{123}), this products become two-valued.
For example,
\begin{equation} \label{IJ1}
IJ_1 = (J_2j_2)J_1 = \begin{cases}J_2(j_2J_1) = J_2J_3 = +
j_1~,\\
-j_2(J_2J_1) =  j_2j_3 =  - j_1 ~.\end{cases}
\end{equation}
From this relation we conclude that we should  introduce negative
sign when removing the brackets and multiply different kinds of
basis units $J^n$ and $j^m$. Then in the first case of the
relation (\ref{IJ1}), because of the product $j_2J_1$, extra minus
arises and both cases give the same result $-j_1$. Using this rule
of opening of brackets, the explicit calculation of the square of
seventh basis unit becomes single valued for any representation
from (\ref{I'}) and we receive $I^2=1$. However, for the
octonionic products for different physical signals this ambiguity
remains and possibly corresponds to the quantum probabilities and
some kind of spin.

Note that, as for ordinary vectors (\ref{e-vector}),
anti-commuting features of octonionic units is not the property of
the basis elements but their binary products. So in the
expressions of the products of $j_n$ with $j_m$ we should not
remove all brackets immediately leading to the mixing of four
vector-like basis units. We should consider binary product of
$j_n$ with one of the vector-like element $J_k$ forming $j_m$ and
then with the second one, that gives the definite result
\begin{equation} \label{jkl}
j_k= -\frac{1}{2}\epsilon_{knm}j^nj^m ~.
\end{equation}
As above we interpret the left product by $j_n$ as the clockwise
and right product as the counter-clockwise rotations. Then the
example of geometrical interpretation of (\ref{jkl}) is
\begin{equation}
j_3j_1 = -j_2 = \setlength{\unitlength}{0.3mm}
\begin{picture}(20, 20)
\thicklines \put(0,5){\line(1,0){20}} \put(20,5){\vector(0,1){10}}
\end{picture}
~\times
\begin{picture}(20, 20)
\thicklines \put(15,0){\line(0,1){17}}
\put(15,17){\vector(-1,-1){11.8}}
\end{picture}
=\frac{\pi}{2}\curvearrowright \begin{picture}(20, 20) \thicklines
\put(15,0){\line(0,1){17}} \put(15,17){\vector(-1,-1){11.8}}
\end{picture}
= \frac{\pi}{2}~
\setlength{\unitlength}{1mm}
\begin{picture}(3,4)\thinlines
\qbezier(2,2)(2,3)(1,3) \qbezier(1,3)(0,3)(0,2)
\put(0,2){\vector(0,-1){3}}
\end{picture}~
\setlength{\unitlength}{0.3mm}
\begin{picture}(20, 20)
\thicklines \put(0,5){\line(1,0){20}} \put(20,5){\vector(0,1){10}}
\end{picture}
~=
\begin{picture}(20, 20)
\thicklines \put(0,10){\line(1,0){20}}
\put(20,10){\vector(-1,-1){11.8}}
\end{picture} ~.
\end{equation}
Similar pictures we have for two other products $j_2j_3 = - j_1$
and $j_1j_2 = - j_3$.

Conjugations, when the vectors $J_n$ change their directions on
the opposite, reverse also the order of elements in any given
expression (physically this means the replacing of causes by the
results)
\begin{eqnarray} \label{tilde}
\widetilde{J_n} = - J_n ~, \nonumber \\
\widetilde{j_n} = \frac{1}{2}\epsilon _{nmk}\widetilde{J^mJ^k} =
\frac{1}{2}\epsilon _{nmk}\widetilde{J^k}\widetilde{J^m} = - j_n~,
\\
\widetilde{I} =\widetilde{J_nj_n} =
\widetilde{j_n}\widetilde{J_n}= - I~.\nonumber
\end{eqnarray}

Finally the formulae (\ref{JjI}), (\ref{OA}), (\ref{j}),
(\ref{Jj}), (\ref{JIjI}), (\ref{jkl}) and (\ref{tilde}) give
multiplication table of all basis elements and thus form the
algebra of split-octonions.

The 'square length' (norm) of a para-vector (\ref{O}), or
(\ref{s}),
\begin{equation} \label{sN}
s^2 = c^2t^2 - x_nx^n  + \hbar^2 \lambda_n\lambda^n -
c^2\hbar^2\omega^2 ~,
\end{equation}
has (4+4)-signature and in general is not positively defined. A
split-octonion does not have an inverse when its norm (\ref{sN})
is zero. To the two critical rotations, in the  planes $(t-x)$, or
$(\lambda-\omega)$ and $(t-\omega )$, or $ (x - \lambda)$, we had
corresponded the two different universal constants $c$ and $\hbar$
in (\ref{O}). It marks an important departure from the Euclidean
space and shows that split-octonions can be used for study of
properties of the real space-time. In the classical limit $\hbar
\rightarrow 0$ the expression (\ref{sN}) reduces to the  ordinary
formula for space-time intervals.

Using the multiplication laws (\ref{JIjI}), a unique mapping of
coordinates of any event into octonion algebra (\ref{O}) can be
equivalently determined as
\begin{equation} \label{s}
s = c(t + \hbar I\omega) + J_n(x^n + \hbar I\lambda^n) ~.
\end{equation}
From this formula we see that pseudo-scalar $I$ introduces the
'quantum' term corresponding to some kind of uncertainty of
space-time coordinates. This terms disappear in the classical
limit $\hbar \rightarrow 0$. Note that physical limits on masses
of reference systems cause uncertainties even for large distances
\cite{NgDa}.

Differentiating (\ref{s}) by the proper time $\tau$ the proper
velocity of a particle can be obtained
\begin{equation} \label{c}
\frac{ds}{d\tau} =  \frac{dt}{d\tau}\left[ c\left( 1 + \hbar
I\frac{d\omega}{dt}\right) + J_n\left(\frac{dx^n}{dt} + \hbar I
\frac{d\lambda^n}{dt}\right) \right] ~.
\end{equation}
From this formula we see that for the critical signals
corresponding to zero norm (\ref{sN}) we have the following
relations
\begin{equation} \label{dtdw}
\frac{\partial}{\partial \lambda^n} = I\hbar
\frac{\partial}{\partial x^n}~, ~~~~~\frac{\partial}{\partial
\omega} = I\hbar \frac{\partial}{\partial t} ~,
\end{equation}
that is similar to introduction of energy and momentum operators
in quantum mechanics.

The invariance of the norm (\ref{sN}) gives the relation
\begin{equation} \label{dtau/dt}
\frac{d\tau}{dt} = \sqrt{ \left[ 1 - \hbar^2
\left(\frac{d\omega}{dt} \right)^2\right] - \frac{v^2}{c^2} \left[
1 - \hbar^2\left(\frac{d\lambda_n}{dx^n}\right)^2\right] } ~,
\end{equation}
where
\begin{equation}
v^2 = \frac{dx_n}{dt}\frac{dx^n}{dt}
\end{equation}
is the velocity measured by the observer (\ref{s}). So the Lorentz
factor (\ref{dtau/dt}) contains extra terms and the dispersion
relation in the (4+4)-space (\ref{sN}) has a form similar to that
of double-special relativity models \cite{double}.

From the requirement to have the positive norm (\ref{sN}) from
(\ref{dtau/dt}) we receive several relations
\begin{equation} \label{delta}
v^2 \leq c ^2 ~, ~~~~~\frac{dx_n}{d \lambda^n} \geq \hbar
~,~~~~~\frac{dt}{d\omega} \geq \hbar ~.
\end{equation}
Recalling that $\lambda$ and $\omega$ have dimensions of
momentum$^{-1}$ and energy$^{-1}$ respectively, we conclude that
uncertainty principle probably has the same geometrical meaning as the
existence of the maximal velocity \cite{Go}.

So some characteristics of physical world (such as dimension,
causality, maximal velocities, quantum behavior, etc.) possibly
connected with the using of normed split algebras, or with our way
of apprehend of reality.

%%%%%%%%%%%%%%%%%%%%%%%%%%%%%%%%%%%%%%%%%%%%%%%%%%%

\section{Conclusion}

In this paper we wanted to introduce some kind of anthrophic
principle in mathematics: some characteristics we usually
attributed to physical world connected with our way to apprehend
reality. To have advantages of division algebras we suggested to
extend vector formalism by embedding it in the algebra of
split-octonions.

Our approach is related with the geometric algebras \cite{He, Ba}
in the sense that we also emphasized the geometric significance of
vector products and avoided matrices and tensors. In distinct from
these models we used non-associative oriented products and tried
to give the physical interpretation to this property.
Non-associativity, which results in difference of left and right
products (what we correspond to causes and effects) could mean
time irreversibility. Also, since the result of product of several
octonions is not single valued, there should appear fundamental
probabilities in calculations of physical processes if they are
done by octonions.

We connected eight real parameters of octonions  with the
space-time coordinates, momentum and energy. To generate complete
basis of split-octonionic the multiplication and distribution laws
of only three vector-like elements are needed, that could result
in our imagination about 3-dimensional character of the space. We
found the Minkowski metric  as the natural metric of octonionic
space in the classical limit $\hbar \rightarrow 0$.

%%%%%%%%%%%%%%%%%%%%%%%%%%%%%%%%%%%%%%%%%%%%%%%%%%%%%%%%%%%%%%%%%%%

\end{document}